# Unconventional Ferroelectric Switching via Local Domain Wall Motion in Multiferroic ε-Fe$_2$O$_3$ Films

*Xiangxiang Guan, Lide Yao*, Konstantin Z. Rushchanskii*, Sampo Inkinen, Richeng Yu*, Marjana Ležaić, Florencio Sánchez, Martí Gich*, and Sebastiaan van Dijken**

X. X. Guan, Prof. R. C. Yu
Beijing National Laboratory of Condensed Matter Physics, Institute of Physics, Chinese Academy of Sciences, Beijing, 100190, China
School of Physical Sciences, University of Chinese Academy of Sciences, Beijing, 10049, China
E-mail: rcyu@aphy.iphy.ac.cn

X. X. Guan, Dr. L. D. Yao, S. Inkinen, Prof. S. van Dijken
NanoSpin, Department of Applied Physics, Aalto University School of Science, P. O. Box 15100, FI-00076 Aalto, Finland
E-mail: lide.yao@aalto.fi, sebastiaan.van.dijken@aalto.fi

Dr. K. Z. Rushchanskii, Dr. M. Ležaić
Peter Grünberg Institut and Institute for Advanced Simulation, Forschungszentrum Jülich and JARA, Jülich 52425, Germany
E-Mail: k.rushchanskii@fz-juelich.de

Dr. F. Sánchez, Dr. M. Gich
Institut de Ciència de Materials de Barcelona (ICMAB), CSIC, Bellaterra, 08193 Catalunya, Spain
E-mail: mgich@icmab.es

Keywords: in-situ STEM, ε-Fe$_2$O$_3$, ferroelectric switching, multiferroics

**Deterministic polarization reversal in ferroelectric and multiferroic films is critical for their exploitation in nanoelectronic devices. While ferroelectricity has been studied for nearly a century, major discrepancies in the reported values of coercive fields and saturation polarization persist in literature for many materials. This raises questions about the atomic-scale mechanisms behind polarization reversal. Unconventional ferroelectric switching in ε-Fe$_2$O$_3$ films, a material that combines ferrimagnetism and ferroelectricity at room temperature, is reported here. High-resolution in-situ scanning transmission electron microscopy (STEM) experiments and first-principles calculations demonstrate that polarization reversal in ε-Fe$_2$O$_3$ occurs around pre-existing domain walls only, triggering local domain wall motion in moderate electric fields of 250 – 500 kV**

**cm$^{-1}$. Calculations indicate that the activation barrier for switching at domain walls is nearly a quarter of that corresponding to the most likely transition paths inside ε-Fe$_2$O$_3$ domains. Moreover, domain walls provide symmetry lowering, which is shown to be necessary for ferroelectric switching. Local polarization reversal in ε-Fe$_2$O$_3$ limits the macroscopic ferroelectric response and offers important hints on how to tailor ferroelectric properties by domain structure design in other relevant ferroelectric materials.**

1. Introduction

A spontaneous polarization which can be switched by applying an electric field is a hallmark of ferroelectrics. While the spontaneous polarization of a crystal is bestowed by its point group symmetry (provided it belongs to the so-called polar classes),[1] its switchability is a far more intricate matter. Indeed, polar materials which cannot be switched, like tourmaline, were already known as pyroelectrics in ancient times because, when heated or cooled, their temperature-dependent spontaneous electric dipole moments give rise to phenomena of electrostatic attraction. In contrast, ferroelectricity was not discovered until 1920,[2] being belated by an essential feature of ferroelectrics; the existence of domains with opposite polarizations, which in as-prepared crystals results in a zero net polarization.[3] The response of ferroelectric domains under applied electric fields determines the characteristics of the polarization versus electric field hysteresis loops which are the quintessence of ferroelectrics and what makes them appealing for most applications. Despite intensive research efforts, we do not have, and might never have, a theory of polarization switching which encompasses all ferroelectrics. Since electric dipoles arise from the relative displacements of positive and negative ions in a crystal and changing their orientation involves a costly deformation of the crystal lattice in terms of elastic energy, the specific structural details can strongly affect the switching mechanism. For instance, many polarization reversal studies have focused on perovskites such as $BaTiO_3$. In this tetragonal structure, the reversal is easily explained by displacements of $Ti^{4+}$ ions above or below the center of its coordination octahedra. However, in $BiFeO_3$, another perovskite with a rhombohedral distortion that has received much attention during the last decade, the polarization switching follows a non-obvious two-step path.[4] To complicate matters further, ferroelectric switching can proceed through the growth of existing domains or with the nucleation and growth of new domains and is strongly affected by the domain geometry or the microstructure of the material.

In this context, it is not surprising that the development path of new ferroelectric oxides beyond the prototypical perovskite family is often not a smooth one. This is particularly true in the case of ferroelectrics for which the energy cost of polarization switching is substantially larger compared to the perovskites, with coercive fields ($E_c$) in the order of MV cm$^{-1}$ instead of tens to hundreds of kV cm$^{-1}$. This is the case for $LiNbO_3$, which was thought to be non-switchable[5] until its ferroelectricity was demonstrated on very thin specimens,[6] and, more recently, doped Hafnia ($Hf_{0.5}Zr_{0.5}O_2$).[7] Another example is $GaFeO_3$, for a long time considered non-ferroelectric until it was prepared in the form of thin films and switched. The fact that $GaFeO_3$ exhibits a $Pna2_1$ polar structure up to high temperatures (above 1350 K)[8] indicates its high stability from which large coercive fields are to be expected. However, most ferroelectricity studies on $GaFeO_3$–type films[9-17] report small to moderate $E_c$ not larger than 500 kV cm$^{-1}$ with a significant dispersion in the values of remanent polarization ($P_r$) ranging from 0.2 µC cm$^{-2}$ [13] to 8 µC cm$^{-2}$.[16] In contrast, Song et al. measured much larger $E_c$ of 1.4 MV cm$^{-1}$ and $P_r$ of 26 µC cm$^{-2}$,[18] in good agreement with theoretical predictions.[18-22]

Multiferroic $\varepsilon$-$Fe_2O_3$ is isostructural to $GaFeO_3$ and has attracted interest in recent years because it combines robust ferrimagnetism and ferroelectricity at room temperature.[23,24] While experiments have demonstrated reversible ferroelectric switching in $\varepsilon$-$Fe_2O_3$ films at room temperature,[24] the atomic-scale mechanism behind polarization reversal is still unclear. Moreover, as in the case of $GaFeO_3$, there are large discrepancies in the reported values of $E_c$, which are as low as 60 kV cm$^{-1}$ (PUND method)[24] and as large as 800 kV cm$^{-1}$ (piezoelectric force microscopy).[16] Here, we provide an alternative explanation for the low $P_r$ values reported on GaFeO$_3$-type ferroelectric films by using in-situ STEM on ε-Fe$_2$O$_3$ films. Our measurements reveal partial switching of the domains at small electric fields which is only localized around pre-existing domain walls. We propose a mechanism of polarization reversal from first-principles calculations. This mechanism highlights the importance of the initial domain structure for the ferroelectric response of GaFeO$_3$-type films and indicates the potential of domain structure design as a tool to lower the polarization reversal barrier or enhance the saturation polarization. In particular, our results can be important to the understanding of relevant ferroelectrics characterized by large coercive electric fields such as LiNbO$_3$ or Hafnia.

## 2. In-Situ STEM Experiments

We performed in-situ STEM experiments on 50-nm-thick ε-Fe$_2$O$_3$ films. The films were grown by pulsed laser deposition (PLD) on an isostructural 8-nm-thick AlFeO$_3$ (AFO) buffer layer and a conducting (111)-oriented Nb-doped SrTiO$_3$ (STO) substrate (see Appendix A for details). We mainly utilized STEM with a high-angle annular dark-field (HAADF) detector. This imaging mode allows recording of incoherent Z-contrast images, in which the brightness of an atomic column is approximately proportional to the square of the average atomic number (Z).[25] Hence, Fe ions are resolved clearly. **Figure** 1 shows HAADF-STEM cross-sectional images and a selected area electron diffraction (SAED) pattern of an ε-Fe$_2$O$_3$ film on our in-situ specimen holder with a piezo-controlled metallic tip. The ε-Fe$_2$O$_3$ and AFO films are epitaxial and their out-of-plane orientation relation with the STO substrate corresponds to (001)$_{ε\text{-Fe2O3}}$//(001)$_{AFO}$//(111)$_{STO}$ (Figure 1b). The polar $c$ axis of the ε-Fe$_2$O$_3$ $Pna2_1$ structure is thus oriented perpendicular to the film plane. The splitting of the film and substrate diffraction spots, as illustrated by the dashed yellow circle in Figure 1b, indicates relaxation of the ε-Fe$_2$O$_3$ lattice with respect to the STO substrate. The microstructure of the ε-Fe$_2$O$_3$ film consists of alternating columnar domains with a width ranging from a few to tens of nanometers (Figure 1c). Fast Fourier transform (FFT) patterns and electron diffraction simulations (Figure S1, Supporting Information) indicate that the domains are oriented along the $\varepsilon$-$Fe_2O_3$ [100] and [110] crystal axes. Diffraction spots of these domains, which we label $D_1$ and $D_2$, are marked by a yellow diamond and red square in the SAED pattern of Figure 1b. The formation of structural domains with approximately equal probability is anticipated because of the pseudo six-fold symmetry in the (004) planes of AFO and $\varepsilon$-$Fe_2O_3$.[24,26]

The unit cell of the $\varepsilon$-$Fe_2O_3$ $Pna2_1$ structure consists of four non-equivalent $Fe^{3+}$ sites, as schematically illustrated in **Figure 2**a for a projection along the [100] crystal axis ($D_1$ domain). The ions labeled Fe1 and Fe2 exhibit a distorted octahedral coordination, while Fe4 ions are tetrahedral. Fe3 ions with regular octahedral coordination are located in the same plane as Fe2 ions, with slightly different coordinates along [001]. Because of this alignment in the [100] projection, it is possible to measure vertical displacements between Fe2 ions and Fe3 ions with high precision. For this reason, we focus our attention on the row with two Fe3 ions in the center and two Fe2 ions at the edges. In the polar $Pna2_1$ phase, the Fe2 ions are positioned either slightly above or below the Fe3 ions, depending on the direction of ferroelectric polarization (see structural models and HAADF-STEM images in Figure 2a). We refer to these two ionic configurations as "smile" and "cry", respectively. Figure 2b shows a high-resolution HAADF-STEM image of a $D_1$ domain in the $\varepsilon$-$Fe_2O_3$ film. In the upper part of the image, the Fe2 and Fe3 ions form a smile structure, whereas cry configurations are measured in the lower part. For clarity, we colored the two polar domains yellow and blue (an uncolored image is shown in Figure S2 of the Supporting Information). In the middle of the image (grey area), the Fe2 and Fe3 ions are positioned on the same horizontal line within measurement error. This alignment suggests that the $\varepsilon$-$Fe_2O_3$ $Pna2_1$ unit cell has transformed into a centrosymmetric structure. Hence, we identify the grey area as a non-polar domain wall. To derive the vertical position of the Fe2 ions with respect to the Fe3 ions inside the polar domains, we used a 2D Gaussian fitting procedure. Averaging over 10 unit cells, we find $d_{Fe2-Fe3} = 6.1 \pm 2.9$ pm for the domain with smiles. A similar value but with opposite polarity is derived for the domain with cries ($d_{Fe2-Fe3} = -8.5 \pm 3.5$ pm).

In order to image vertical displacements between Fe2 and Fe3 ions under an electric field, we contacted the ε-Fe$_2$O$_3$ film by the piezo-controlled metallic tip of our in-situ probing holder (Figure 1a). In the in-situ measurements, we grounded the tip and applied triangular voltage pulses with a duration of 0.1 s to a contact on the Nb-doped STO substrate (see Methods Section for details). The pulse amplitude was swept and high-resolution images were recorded after applying voltages with opposite polarity. From a careful analysis of the HAADF-STEM images, we conclude that the Fe2 and Fe3 ions do not measurably shift inside the domains, i.e., away from the domain wall, up to a pulse amplitude of 5 volts. In the immediate vicinity of the wall, however, we observe reversible transpositions of Fe ions, turning smiles into cries and vice versa. **Figure 3**a shows snapshots from this experiment (full images are shown in Figure S3 of the Supporting Information). We focus on one unit cell above and another below the domain wall (marked by red and cyan rectangles in Figure 2b) and plot the relative positions of the two Fe2 ions with respect to the mean position of the two Fe3 ions (i.e. $d_{Fe2-Fe3}$). Applying a +3 V pulse turns the initial smile into a cry, while the initial cry remains practically unaltered. In contrast, negative voltage pulses turn cries into smiles. Since positive (negative) pulses move the positively charged Fe$^{3+}$ ions up (down), our in-situ results imply that the Fe3 ions move more than the Fe2 ions in an applied electric field. Moreover, from the bias polarity we derive that the ferroelectric polarization points up in the cry domain and down in the smile domain. The relative vertical displacement of the Fe2 and Fe3 ions near the domain boundary is consistent with the magnitude of smiles and cries inside the domains. This observation suggests that applying 1.5 – 3.0 V, corresponding to an electric field of 250 – 500 kV cm$^{-1}$, is sufficient to saturate the local polarization, in agreement with the polarization versus electric field hysteresis loop previously reported for this film.[24]

At higher bias voltages, local polarization reversal leads to lateral motion of the domain wall. We illustrate this effect in **Figure 4**. The depicted HAADF-STEM images show how the cry domain with upward polarization grows at the expense of the smile domain when we apply positive voltage pulses of increasing amplitude to the Nb-doped STO substrate. This behavior, driven by local polarization reversal at the domain wall, produces a macroscopic ferroelectric response. Since ferroelectric switching does not encompass the entire sample (Figure S4, Supporting Information), we expect the magnitude of the macroscopic signal to be small compared to predictions based on first-principles calculations. The restriction of polarization reversal to the vicinity of pre-existing domain walls might explain existing discrepancies between theory and experiments on $\varepsilon$-Fe$_2$O$_3$ and possibly GaFeO$_3$ in literature.

## 3. First-Principles Calculations

### A. Ferroelectric Switching *without* Pre-Existing Domain Walls

Recent studies on ferroelectric switching in isostructural GaFeO$_3$ considered the *Pnna* phase as non-polar reference structure, yielding high energy barriers for ionic displacements.[18-22] Our in-situ STEM experiments on $\varepsilon$-Fe$_2$O$_3$ films, however, indicate that ferroelectric switching occurs near domain walls at moderate electric fields. In order to understand this low-energy switching behavior, we performed an *ab initio*-based search for alternative transition paths using the USPEX code.[27-30] Based on this analysis, we found a low-energy metastable centrosymmetric phase with monoclinic *P*2/*c* symmetry (**Table 1**). The *P*2/*c* phase is an elastically relaxed extension of the *Pbcn* structure (the latter was recently obtained by K. Xu *et al*. using a stochastic surface walking method[31]) with frozen distortions according to the $\Gamma_4^+$ irreducible representation of *Pbcn* (**Figure 5**a–d). We estimate the potential barrier for ferroelectric switching via the *Pbcn* reference structure at 86 meV/f.u. (similar to 85 meV/f.u. obtained in Ref. 31), whereas ferroelectric switching inside a $\varepsilon$-Fe$_2$O$_3$ domain via the elastically relaxed *P*2/*c* structure results in a significantly lower energy barrier of 59 meV/f.u. (Figure 5e).

An examination of the Γ-point phonon structure reveals a dynamic instability in the *Pbcn* phase, with one polar phonon of imaginary frequency. Figures 5f and g show the eigendisplacements of Fe ions for the unstable phonon mode in *Pbcn*. Clearly, the eigendisplacements produce smile features similar to those in the STEM images of Figure 2–4, with Fe3 ions shifting down and Fe2 ions shifting up along the [001] direction. Freezing of the unstable polar mode lowers the symmetry of the reference structure from non-polar *Pbcn* to polar *Pna*$2_1$. A decomposition of real ion displacements during these structural conversions reveals the main contribution of the low energy $\Gamma_3^-$ phonon, i.e., its eigenvector completely defines the transition path for polarization switching. Although the phonon develops some polarization along the [001] direction in the reference non-polar structures (Figure 5f), ions mainly displace along the [100] direction, producing large antipolar components in neighboring chains (running along [001]) of Fe ions (Figure 5g). Consequently, polarization switching along [001] is accompanied by significant antipolar displacements of a large number of ions along [100] (see Figure 5h). Since the displacements of Fe ions along the in-plane directions are not stabilized against thermal fluctuations by a perpendicular electric field, this fact could explain why switching within the ε-$Fe_2O_3$ domains is not observed in our in-situ STEM experiments.

In contrast to *Pbcn*, all Γ-phonons are real in the *P*2/*c* structure. The eigenvector of the low-energy polar phonon mode in the *P*2/*c* phase exhibits all the main features of the imaginary mode vector of *Pbcn*. Consequently, the eigendisplacements of Fe ions are comparable for both structures. Decomposition of the real polar displacements in terms of eigendisplacements of the non-polar reference structure reveals the principal contribution of the polar $\Gamma_2^-$ mode together with the full-symmetry $\Gamma_1^+$ mode (Figure 5d). Coupling of these two modes produces a triple-well potential (Figure 5e), similar to the one in monoclinic $Sn_2P_2S_6$ uniaxial ferroelectrics.[32] In the *P*2/*c* structure, the Fe1 sublattice of the *Pbcn* structure with tetrahedral coordination splits into two sublattices, one being tetrahedrally and the other octahedrally coordinated by oxygen ions (Figure 5a). Transformation to a polar phase from *P*2/*c* is therefore simpler than from *Pbcn* because it involves asymmetric ion displacements in only half of the unit cells. To develop a smile feature starting from the *P*2/*c* structure (indicated by purple spheres in Figure 5i), the largest displacements are performed by Fe ions with positions $z \geq 0.5$ (compare distances between purple and yellow spheres), whereas Fe ions with $z \leq 0.5$ displace most during the formation of a cry feature (compare distances between purple and blue spheres). This displacement pattern is consistent with the eigendisplacements of the phonon mode with lowest energy. As a result, the non-polar *P*2/*c* structure transforms to the polar *Pc* structure. However, because elastic relaxations decrease the monoclinic angle to zero, the system finally settles in the *Pna*2$_1$ orthorhombic polar phase. Therefore, in order to switch the polarization via the *P*2/*c* reference structure one has to break the symmetry in such a way that the angle between the *a* and *c* axes of *Pna*2$_1$ becomes monoclinic, which requires an ε$_5$ shear strain component.

In summary, one can formulate several reasons why ferroelectric switching inside ε-Fe$_2$O$_3$ domains is not observed in electric fields of up to 2 MV cm$^{-1}$ (maximum electric field in our in-situ STEM experiments): (i) The low-energy path for switching goes via the non-polar *P*2/*c* state. Since this transition state has lower symmetry then the initial *Pna*2$_1$ polar phase, the symmetry of *Pna*2$_1$ needs to be lowered during switching. In particular, this requires breaking of the glide plane symmetry orthogonal to [100]. This could be achieved by an elastic deformation ε$_5$ via the piezoelectric tensor component d$_{51}$, which we found to be significant in ε-Fe$_2$O$_3$ (see **Table 2**). However, in this scenario the electric field should have a [100] in-plane component, which is not the case for the out-of-plane geometry used in our experiment. (ii) One can consider *Pbcn* as a virtual centrosymmetric reference structure with a higher transition barrier than *P*2/*c*, but also in this scenario the main component of the phonon mode that drives the transition lies along [100], i.e., perpendicular to the applied electric field. The out-of-phase ion displacements along [100] are fully symmetric due to glide plane symmetry. Hence, one can consider ferroelectric switching in ε-Fe$_2$O$_3$ as a side-product of antiferroelectric ion displacements along the [100] axis. Reversal of the displacements along [100] will reverse also the net polarization along [001]. The estimated ratio of Fe ion displacements along [100] and [001] is about ~3.5 for Fe1 and Fe4, ~4.2 for Fe3, and ~8.7 for Fe2 ions. For oxygen ions the corresponding ratio varies from ~2.4 to ~23. These numbers suggest that only a very high out-of-plane electric field component would produce the corresponding in-plane displacements, which could be well beyond dielectric breakdown of the material. Again, also here an in-plane electric field component along the [100] direction would assist in ferroelectric switching.

**B. Ferroelectric Switching *with* Pre-Existing Domain Walls**

Next, we discuss the role of domain walls in ferroelectric switching. The metastable non-polar *P*2/*c* structure could exist as an extended phase or appear as a non-polar domain wall between ferroelectric domains with opposite polarizations. As can be seen from the triple-well potential for switching via the *P*2/*c* path (Figure 5e), the barrier from the non-polar to the polar phase is very small (~1 meV/f.u.). Hence, the *P*2/*c* state is not stable in its free-standing form. However, the structure could be realized between head-to-head or tail-to-tail domains, as observed experimentally in the STEM image of Figure 2b (domain stacking along the [001] direction). Because of its significant monoclinic distortion (see Table 1), it is unlikely that the *P*2/*c* phase separates orthorhombic domains that are stacked along the [010] direction and polarized along [001]. In this geometry, which we observe also in experiments (Figure 4), we expect the domain wall to have an orthorhombic structure, probably with lower symmetry than *Pbcn*.

To analyze the effect of domain walls on ferroelectric switching, we consider a vertical wall separating two polar *Pna*2$_1$ domains wherein the polarization points down (smiles) and up (cries), respectively (**Figure 6**a). In contrast to switching via *Pbcn* and *P*2/*c* structures inside a ε-Fe$_2$O$_3$ domain where Fe1 and Fe4 ions occupy octahedral and tetrahedral sites in the initial *Pna*2$_1$ phase, respectively (Figure 5c), the unit cell of the modeled domain wall is altered by cross-substituting one tetrahedrally coordinated Fe4 ion with an octahedral Fe ion and one octahedrally coordinated Fe1 ion with a tetrahedral Fe ion (see blue ions in Figure 6a). These substitutions break the glide plane symmetry orthogonal to the [100] axis, symmetrize the hexagon formed by the Fe2-Fe4-Fe4-Fe2-Fe1-Fe1 ions, and align the Fe2 and Fe3 ions along the [001] directions within the domain wall. HAADF-STEM simulations of this domain wall configuration shown in Figure S5 of the Supporting Information indicate that smile and cry features are indeed not expected to appear in the domain wall, in agreement with the STEM images of Figures 2 and 4. By moving the domain as a whole unit along the [010] direction, we estimate an energy barrier of ~22 meV/f.u., which is significantly lower than that estimated for ferroelectric switching inside a ε-$Fe_2O_3$ domain (Figure 5e). Detailed analysis of the calculations shows that one of the octahedral Fe1 ions swaps to a tetrahedral surrounding and one of the tetrahedral Fe4 ions swaps to an octahedral surrounding during the first stages of lateral domain wall translation (Figure 6a–f). Next, the second Fe1 and Fe4 ions undergo the same coordination modifications. Coincidently, the cry feature transforms into a smile feature via an intermediate phase by vertical transpositions of Fe2 and Fe3 ions if the domain wall moves from left to right. This observation is in excellent agreement with our in-situ STEM measurements of voltage-induced motion of non-polar domain walls (Figure 4). Figure 6g shows the displacements of Fe ions along the [100] direction across a domain wall. The Fe2 and Fe3 ions displace gradually in the vicinity of the wall, undergoing only small transpositions upon domain wall motion. From the eight Fe ions taking part in ferroelectric switching via lateral wall motion, the ions on the Fe1 and Fe4 sites displace most. Their displacement is ~0.4 Å, which is only half compared to switching inside a ε-$Fe_2O_3$ domain. Consequently, the energy barrier of domain wall assisted switching is significantly smaller than that of switching inside ε-$Fe_2O_3$ domains if the walls pre-exist. If the domain walls need to form first, an extra price has to be paid. For a discussion on the influence of lattice deformations on the energy barrier of ferroelectric switching, we refer to Figures S6 and S7 of the Supporting Information.

In addition to the vertical domain wall, we considered an inclined domain wall (Figure S8, Supporting Information), which is also visible in the STEM images (see Figure 4). As in the vertical wall, the domains are stacked along the [010] direction, but the inclined wall is not orthogonal to [010]. We find that in contrast to the vertical domain wall, the inclined domain wall preserves the orthogonal to [100] glide plane symmetry. Therefore, taking into account the above mentioned arguments, we do not expect that this wall plays a role in polarization switching. Unfortunately, the estimated formation energy of the inclined wall is significantly smaller than that of the vertical wall, $E_f^{\text{inclined}} = 234$ mJ m$^{-2}$ versus $E_f^{\text{vertical}} = 370$ mJ m$^{-2}$. This could lead to a transformation of vertical walls into inclined walls during switching, which would result in the disappearance of the ferroelectric response. In both cases, the formation energy is much higher than in the conventional ferroelectrics (e.g. ~10 mJ m$^{-2}$ in BaTiO$_3$ [33]). It is even higher than the most energy expensive 180° domain wall in PbTiO$_3$ (~169 mJ m$^{-2}$ [33]). This makes the nucleation of domains with opposite polarization very difficult and explains the lack of ferroelectric switching inside ε-Fe$_2$O$_3$ domains. To generate a ferroelectric response in ε-Fe$_2$O$_3$, one has to grow thin films with pre-engineered domain walls.

## 4. Conclusions

We combined in-situ STEM experiments and first-principles calculations to elucidate the origin of ferroelectric switching in multiferroic ε-Fe$_2$O$_3$ films. We find that ferroelectric switching proceeds locally via the motion of non-polar boundaries. During switching, the polar state of the *Pna*2$_1$ structure transforms via an intermediate non-polar *P*2/*c* or *Pbcn* structure. Since the energy barrier of this phase transition is reduced at domain boundaries, we experimentally measure a local ferroelectric response at moderate electric fields of 250 – 500 kV cm$^{-2}$. Ferroelectric switching by a perpendicular electric field is prevented inside domains because the activation barrier is larger and it involves significant antipolar ionic displacements within the film plane. The local nature of polarization reversal provides a viable explanation for discrepancies in the reported values of coercive field and saturation magnetization for ε-Fe$_2$O$_3$ and isostructural GaFeO$_3$. We argue that similar atomic-scale mechanisms may also be at play in other relevant ferroelectric materials, especially those with large activation barriers.

## 5. Methods Section

*Film Growth*: ε-AlFeO$_3$/ε-Fe$_2$O$_3$ films were grown by pulsed laser deposition (PLD) using a KrF excimer laser (248 nm wavelength, 25 ns pulse duration) at 5 Hz repetition rate on Nb-doped (0.5%) (111)-oriented SrTiO$_3$ (STO) single crystals at 825ºC, as described in more detail in Ref. 24. For the present study, we employed the film on which ferroelectric characterization reported in Ref. 24 was previously studied.

*In-Situ STEM Experiments*: High-resolution in-situ imaging of ion displacements in ε-Fe$_2$O$_3$ thin films was carried out on an aberration-corrected JEOL 2200FS TEM using a piezo-controlled in-situ biasing holder (HE150 electrical probing holder with double-tilt capability from Nanofactory AB). Wedge-shaped TEM specimens were prepared by a mechanical polishing process and argon ion milling (for details see Ref. 34). We used silver paste for making good electrical contacts to the Nb-doped STO substrate after gluing the specimen to a half Cu TEM grid. As top electrode, we exploited the piezo-controlled Pt/Ir tip of the probing holder (Figure 1a). This tip contacted the sample over a length of about 30 nm. During in-situ STEM measurements, the tip was grounded and triangular voltage pulses were applied to the Nb-doped STO substrate. The duration of the voltage pulses was set to 0.1 s and their magnitude was continuously swept in 0.1 V steps.

*First-Principles Calculations*: For structural predictions, we employed the USPEX code,[27-30] which is based on an evolutionary algorithm. The USPEX method searches for the global ground state structure, but produces a series of metastable states as by-product, being therefore suited also for searching all possible metastable polymorphs. We applied this method to search for the low-energy metastable centrosymmetric structure of ε-Fe$_2$O$_3$. We considered a 40-atoms unit cell of ε-Fe$_2$O$_3$. Each individual structure was relaxed by means of the Vienna ab-initio simulation package VASP,[35-37] with projector augmented wave pseudopotentials.[38] We considered the following valence-electron configurations: $3p^63d^74s^1$ for Fe and $2s^22p^4$ for oxygen. Ground-state properties were obtained using the generalized gradient approximation (PBEsol) adopted for solids.[39]

To correctly account for strong electron correlation effects on the d shell of Fe atoms, a DFT+U scheme was used.[40] We used on-site Coulomb parameter $U$ = 4.0 eV and Hund's exchange $J$ = 0.89 eV. The kinetic energy cutoff was set to 600 eV. For Brillouin zone sampling, we used uniform Γ-centered meshes with a reciprocal-space resolution of $2\pi \times 0.05$/ Å. All energies were converged to within $10^{-6}$ eV per cell. The stopping criterion for structural relaxation was a change in (free) energy smaller than 1 meV.

Searches for metastable structures with the USPEX code were made by assuming ferromagnetic ordering on all Fe atoms in order to make energy comparisons of different structures independent of possible magnetic configurations. Total energy for the resulting low-energy structures as well as phonon calculations were performed with proper antiferromagnetic alignments of the spins.

For phonon calculations, the obtained structures were optimized with residual forces below 0.002 eV/Å. The Brillouin zone was sampled with $4 \times 4 \times 4$ Monkhorst-Pack Γ-centered special k-point grids for *P*2/*c* and *Pbcn* structures. Calculations of dynamical properties were performed using the force-constant method.[41] Eigenvalues and eigenvectors of the long-wavelength phonons were calculated at the Γ-point of the Brillouin zone and, therefore, no supercells were used. The Hellman-Feynman forces were calculated for displacements of atoms of up to 0.04 Å. The dynamical matrix was constructed by Fourier transforming the force constant calculated at the Γ-point. The phonon-mode frequencies and atomic displacement pattern were obtained as eigenvalues and eigenvectors of the dynamical matrix.

Domain wall movements were simulated in 320-unit cell, which consists of two equivalent domains with opposite direction of polarization. Ferromagnetic ordering on all Fe atoms was imposed for initial and final states as well as for all images along the transition path to avoid possible uncontrolled flips of spins on individual atoms during nudged elastic band calculations.

Transition energy barriers were calculated with a variable cell nudged elastic band as implemented in USPEX code.[42] We used crystallographic codes AMPLIMODE[43] and FINDSYM[44] for symmetry analysis of the obtained phases and their relations, as well as VESTA code[45] for visualization of obtained results.

**Acknowledgements**

This work was supported by the Academy of Finland (Grant Nos. 293929, 304291, 319218 and 316857) and by the European Research Council (ERC-2012-StG 307502 and ERC-2018-CoG 819623). STEM analysis was conducted at the Aalto University OtaNano-Nanomicroscopy Center (Aalto-NMC). ICMAB acknowledges financial support from the Spanish Ministry of Economy, Competitiveness and Universities, through the "Severo Ochoa" Programme for Centres of Excellence in R&D (SEV- 2015-0496) and the project MAT2017-85232-R co-financed with the EU FEDER program, as well as the Generalitat de Catalunya (projects 2017SGR1377 and 2017SGR765). We thank Jaume Gàzquez for discussions about our work and critical reading of the manuscript. K.Z.R. and M.L. gratefully acknowledge the financial support by Deutsche Forschungsgemeinschaft (DFG) through DFG-ANR GALIMEO project (LE 2504/2-1), as well as the support of Jülich Supercomputing Centre (JSC, project jiff38) and JARA-HPC Partition (projects jara0081 and jara0126). R.C.Y and X.X.G acknowledge the financial support by the National Key Research Program of China (Grant No. 2017YFA0206200) and the National Natural Science Foundation of China (Grant No. 11874413).

**Figures**

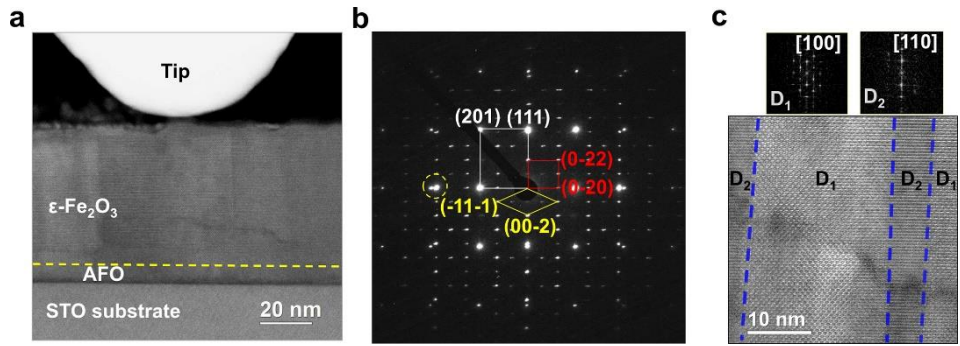

**Figure 1.** In-situ STEM measurement geometry and structural characterization. a) Low magnification cross-sectional HAADF-STEM image of a 50 nm ε-Fe$_2$O$_3$/8 nm AFO/Nb-doped STO(111) specimen after establishing contact with a Pt/Ir tip of the piezo-controlled probing holder. The dashed yellow line marks the interface between the ε-Fe$_2$O$_3$ and AFO films. During in-situ measurements, the Pt/Ir tip is grounded and voltage pulses are applied to the conducting STO substrate. b) Selected area electron diffraction (SAED) pattern of the sample. The pattern is indexed by considering a superposition of diffraction peaks along the zone axis [$\bar{1}\bar{1}2$] of the STO substrate (white rectangle) and [100] and [110] of the ε-Fe$_2$O$_3$ film (red rectangle and yellow diamond). Splitting of the ε-Fe$_2$O$_3$ and STO diffraction spots (see yellow circle) confirms relaxation of the ε-Fe$_2$O$_3$ film. c) Enlarged area of the ε-Fe$_2$O$_3$ thin film showing the presence of alternating D$_1$ and D$_2$ structural domains. The panels above the image show corresponding FFT patterns.

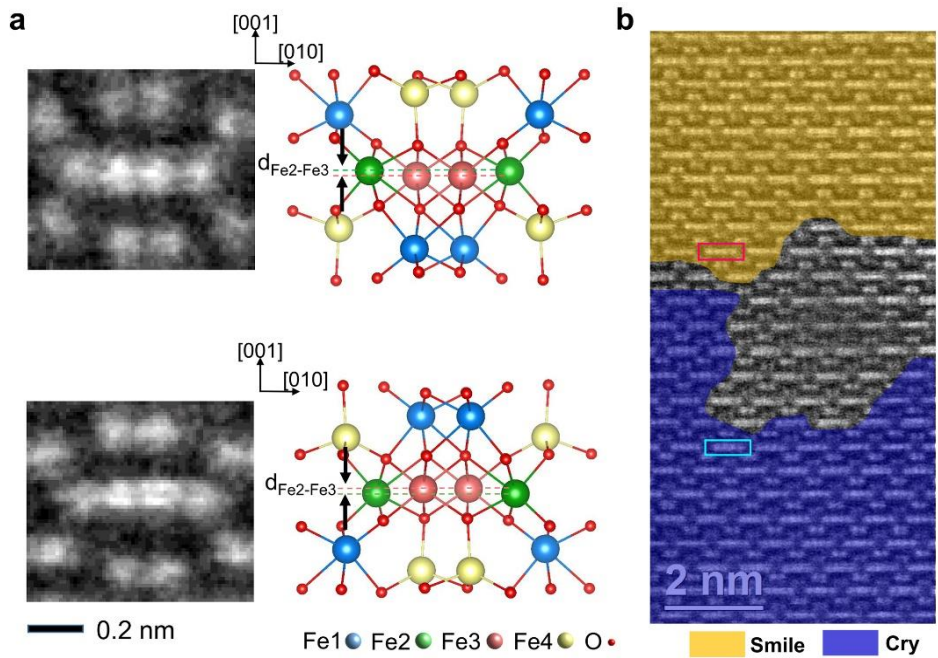

**Figure 2**. Ferroelectric domain structure. a) High-resolution HAADF-STEM images along the [100] zone axis of the ε-$Fe_2O_3$ film showing a smile (upper panel) and cry (lower panel) alignment of the Fe2 and Fe3 ions. The corresponding structural models are shown on the right. b) HAADF-STEM image of two polar domains in the ε-$Fe_2O_3$ film (yellow and blue areas) separated by a non-polar boundary (grey area). The red and cyan rectangles mark Fe ions whose relative positions are monitored during the application of electric field pulses (see Figure 3 for results).

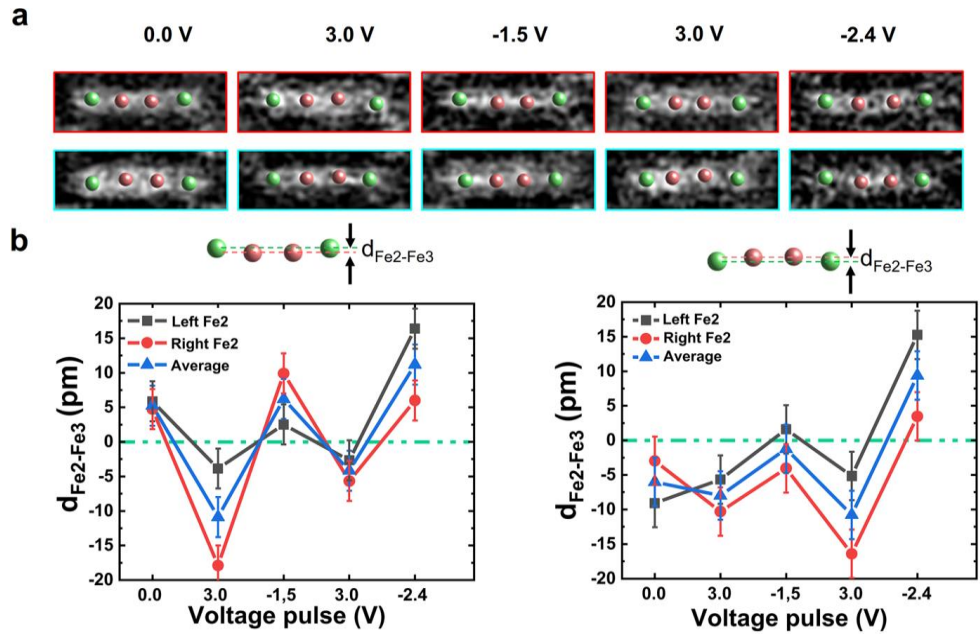

**Figure 3.** Electric-field-driven ion displacements. a) High-resolution HAADF-STEM images from selected areas in Figure 2b obtained after application of alternating positive and negative voltage pulses to the Nb-doped STO substrate. The positions of the Fe2 and Fe3 ions are determined by a Gaussian fitting procedure and marked by green and red dots, respectively. b) Summary of the relative vertical positions of the left Fe2 ion (grey data) and right Fe2 ion (red data) with respect to the average vertical position of the Fe3 ions after the application of positive and negative voltage pulses. The average $d_{Fe2-Fe3}$ displacement is shown in blue. The left and right graphs show results for the red and cyan rectangles in Figure 2b. Positive values of $d_{Fe2-Fe3}$ correspond to a smile, while negative values resemble a cry.

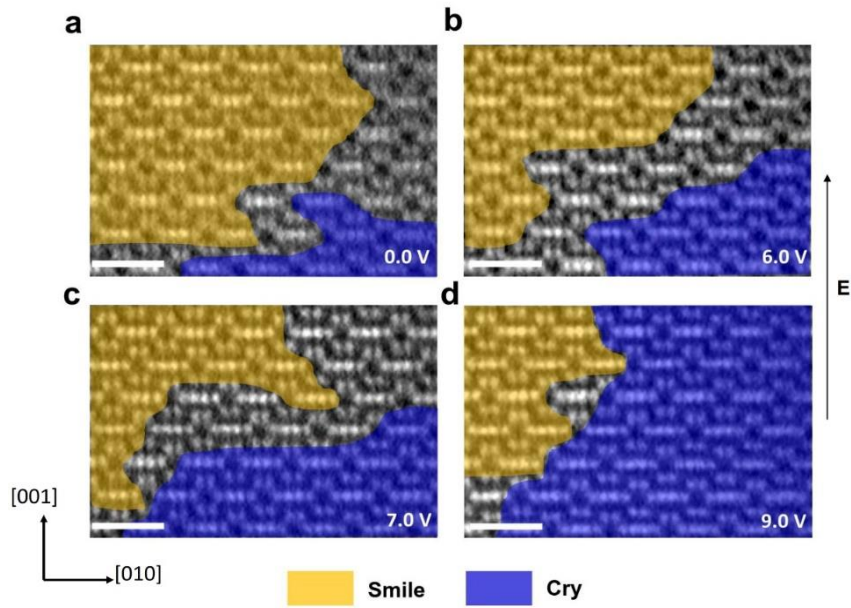

**Figure 4**. In-situ STEM observation of the motion of a ferroelectric domain boundary. a)-d) HAADF-STEM images showing the evolution of two polar domains during application of positive voltage pulses to the Nb-doped STO substrate. In the area with cries (blue) the polarization points up and in the area with smiles (yellow) it points down. Local polarization reversal near the non-polar boundary leads to domain wall motion. The scale bar corresponds to 1 nm.

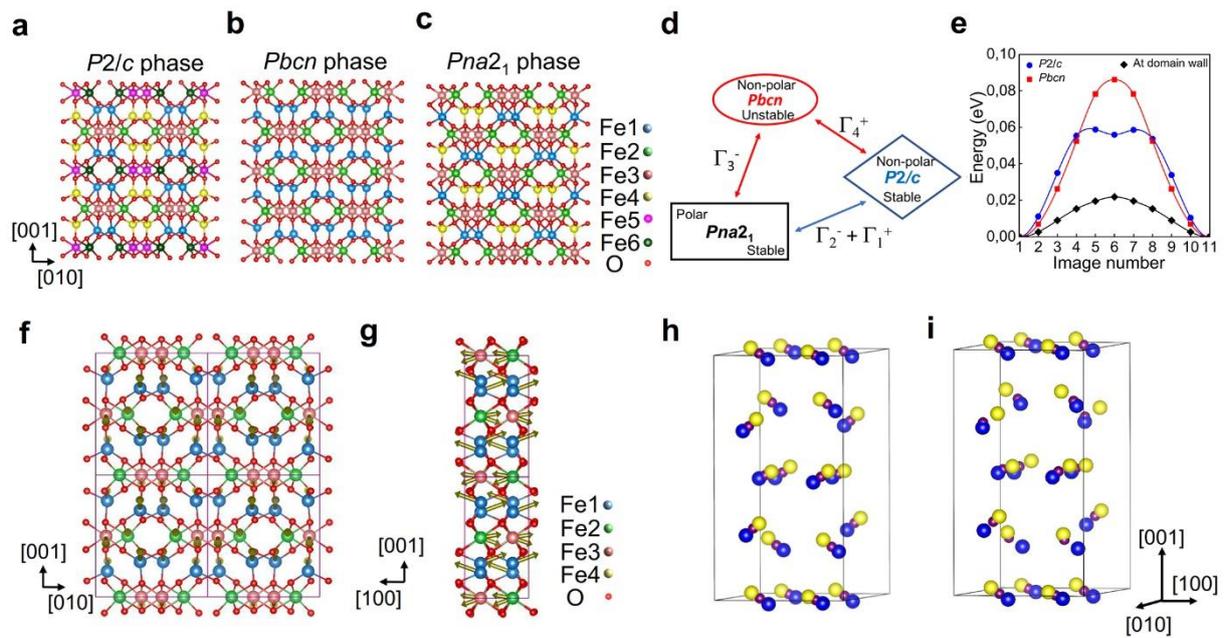

**Figure 5.** Ferroelectric switching inside ε-Fe$_2$O$_3$ domains. a)-c) Structural models of the *P2/c*, *Pbcn*, and *Pna2$_1$* phases of ε-Fe$_2$O$_3$. d) Relations between the polar ground state and non-polar transition states of ε-Fe$_2$O$_3$. e) Calculated migration barriers for switching via *P2/c* and *Pbcn* reference structures inside a ε-Fe$_2$O$_3$ domain, as well as in the presence of a domain wall. Image numbers indicate intermediate images on the transition path between the initial and final states. f),g) Eigendisplacements of Fe ions in the imaginary soft-mode of the non-polar *Pbcn* structure along (a) [100] and (b) [010]. The eigendisplacements of Fe ions in the low-energy phonon mode of the elastically relaxed *P2/c* structure are comparable to those shown here. h),i) Schemes of ferroelectric switching via (h) *Pbcn* and (i) *P2/c*. The view is along the [010] axis and, for clarity, only Fe ions are shown. The images depict two states with opposite polarization directions with blue and yellow spheres indicating ion positions in the states corresponding to the cry and smile features, respectively. The purple spheres represent the positions of ions in the transition state. The calculations show that all Fe ions move equivalently if switching proceeds though the *Pbcn* phase. In this case, the ions shift by ~0.8 Å. For switching via the *P2/c* path in (i), the transposition of Fe ions is clearly asymmetric. Ions in the transition state are close to the positions corresponding to the cry features (blue spheres) in the upper half of the unit cell, whereas they are close to the positions corresponding to the smile features (yellow spheres) in the lower part of the picture. This means that only a part of the Fe ions displace during the first half of the transition and the remaining ions finalize the transition during the second half. The absolute displacement of Fe ions during switching via the *P2/c* path is similar to the *Pbcn* case, i.e., ~0.8 Å.

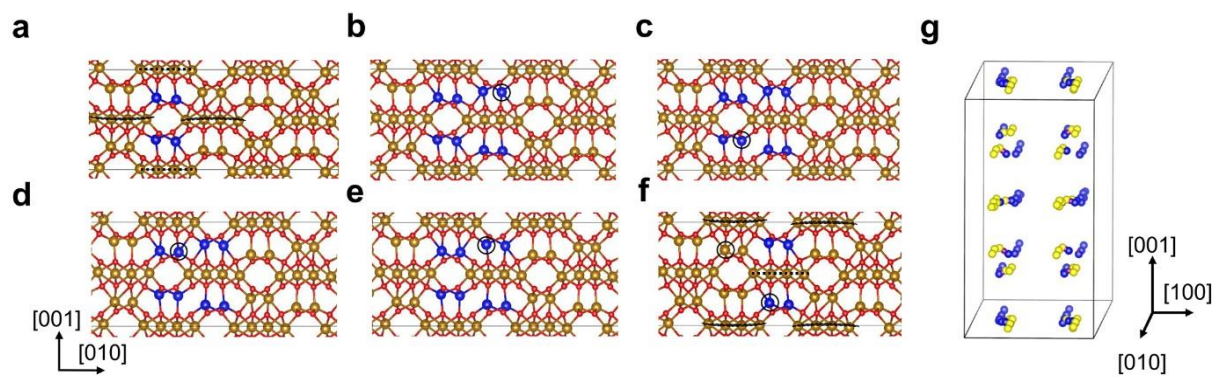

**Figure 6.** a) Model of a 180° domain wall separating two polar $Pna2_1$ domains with opposite polarization. The domains are stacked along the [010] direction. b)-f) Calculated structure sequence during local ferroelectric switching at the domain wall. Switching involves the swapping of the Fe1 and Fe4 sublattices, which is achieved by lateral motion of the non-polar domain wall. The blue circles indicate ions whose coordination changed during various stages of the switching process. g) Displacement of Fe ions during domain wall assisted ferroelectric switching. Fe ion positions in the cry, smile, and transition states are indicated by blue, yellow, and purple spheres, respectively. The maximum ion displacement during switching is ~0.4 Å.

| Space group | Lattice constant, Å | Wyckoff position | x | y | z |
|---|---|---|---|---|---|
| P2/c | a= 9.60108 | Fe1 4g | 0.21776 | 0.09929 | 0.28643 |
| | | Fe2 2f | 0.5 | 0.27769 | 0.25 |
| | | Fe3 2f | 0.5 | -0.08640 | 0.25 |
| | b= 8.71696 | Fe4 4g | 0.29853 | 0.39862 | 0.70943 |
| | | Fe5 2e | 0.0 | 0.78847 | 0.25 |
| | c= 5.04469 | Fe6 2e | 0.0 | 0.40230 | 0.25 |
| | β= 96.32° | | | | |
| | | O1 4g | 0.61657 | -0.08336 | -0.06784 |
| | | O2 4g | 0.89069 | 0.58735 | 0.39074 |
| | | O3 4g | 0.62095 | 0.76342 | 0.46437 |
| | | O4 4g | 0.88562 | 0.74948 | -0.09903 |
| | | O5 4g | 0.86415 | 0.08207 | 0.88130 |
| | | O6 4g | 0.62675 | 0.41975 | 0.43261 |
| | | | | | |
| Pbcn | a= 9.59875 | Fe1 4c | 0.0 | 0.09203 | 0.25 |
| | b= 8.72242 | Fe2 8d | 0.21110 | 0.39648 | 0.24759 |
| | c= 5.00751 | Fe3 4c | 0.0 | 0.71392 | 0.25 |
| | | O1 8d | 0.61160 | 0.58726 | 0.58818 |
| | | O2 8d | 0.61792 | 0.74270 | 0.06735 |
| | | O3 8d | 0.86843 | 0.42188 | 0.59656 |

**Table 1.** Structural parameters of non-polar monoclinic *P*2/*c* and orthorhombic *Pbcn* reference structures

| $d_{ij}$ | 1 | 2 | 3 | 4 | 5 | 6 |
|---|---|---|---|---|---|---|
| 1 | | | | | 7.84 | |
| 2 | | | | 1.31 | | |
| 3 | -0.178 | -3.18 | 4.23 | | | |

**Table 2.** Calculated piezoelectric tensor (in $10^{-12}$ C/N) of ε-$Fe_2O_3$ in *Pna*$2_1$ crystallographic setting.

**Unconventional Ferroelectric Switching via Local Domain Wall Motion in Multiferroic ε-Fe$_2$O$_3$ Films**

*Xiangxiang Guan, Lide Yao[*], Konstantin Z. Rushchanskii[*], Sampo Inkinen, Richeng Yu[*], Marjana Ležaić, Florencio Sánchez, Martí Gich[*], and Sebastiaan van Dijken[*]*

X. X. Guan, Prof. R. C. Yu
Beijing National Laboratory of Condensed Matter Physics, Institute of Physics, Chinese Academy of Sciences, Beijing, 100190, China
School of Physical Sciences, University of Chinese Academy of Sciences, Beijing, 10049, China
E-mail: rcyu@aphy.iphy.ac.cn

X. X. Guan, Dr. L. D. Yao, S. Inkinen, Prof. S. van Dijken
NanoSpin, Department of Applied Physics, Aalto University School of Science, P. O. Box 15100, FI-00076 Aalto, Finland
E-mail: lide.yao@aalto.fi, sebastiaan.van.dijken@aalto.fi

Dr. K. Z. Rushchanskii, Dr. M. Ležaić
Peter Grünberg Institut and Institute for Advanced Simulation, Forschungszentrum Jülich and JARA, Jülich 52425, Germany
E-Mail: k.rushchanskii@fz-juelich.de

Dr. F. Sánchez, Dr. M. Gich
Institut de Ciència de Materials de Barcelona (ICMAB), CSIC, Bellaterra, 08193 Catalunya, Spain
E-mail: mgich@icmab.es

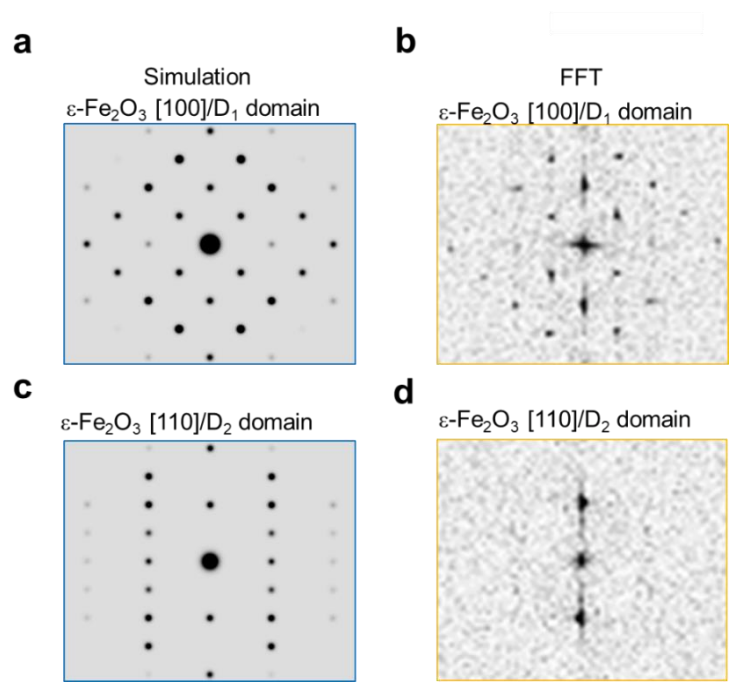

**Figure S1.** Electron diffraction simulations and fast Fourier transform patterns of experimental HAADF-STEM images along the ε-Fe$_2$O$_3$ [100] (a,b) and [110] (c,d) crystal axes. The data in (a,b) and (c,d) correspond to D$_1$ and D$_2$ structural domains, respectively.

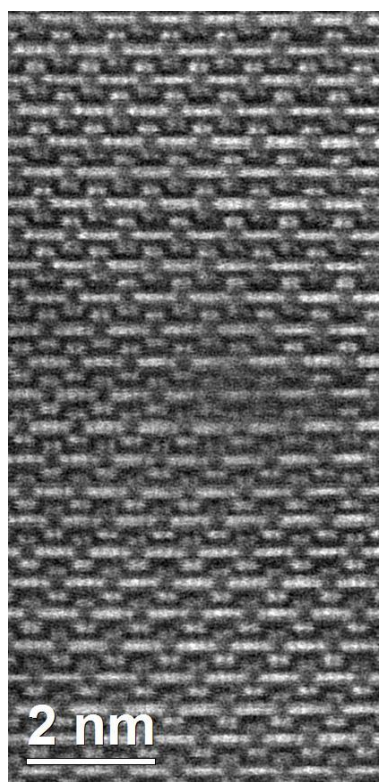

**Figure S2.** High-resolution HAADF-STEM image of a D$_1$ domain in the ε-Fe$_2$O$_3$ film. This is an uncolored version of Figure 2b in the main manuscript.

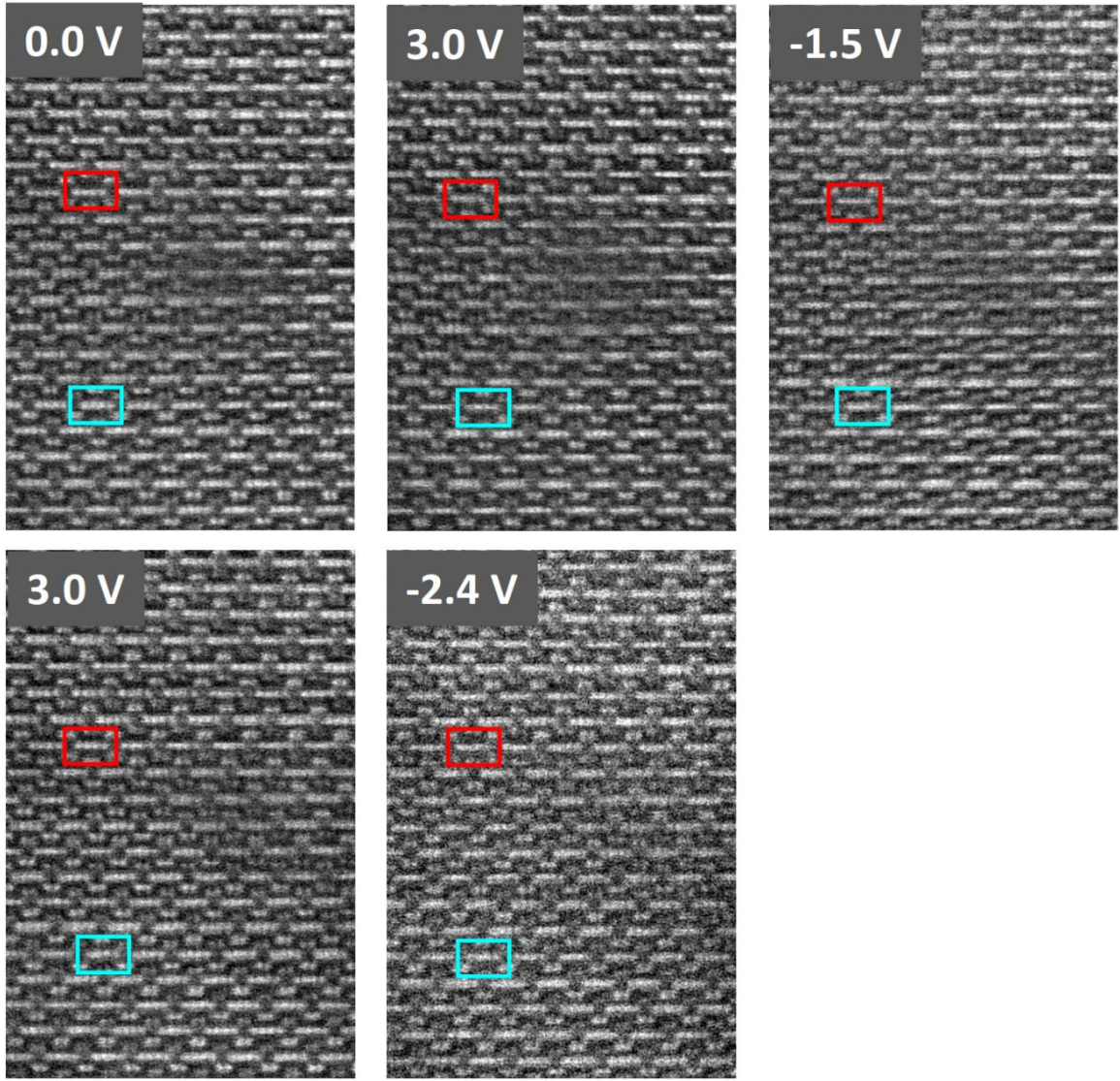

**Figure S3**. High-resolution HAADF-STEM images of the ε-Fe$_2$O$_3$ film obtained after application of alternating positive and negative voltage pulses to the Nb-doped STO substrate. The data of Figure 3 in the main manuscript are extracted from the areas indicated by red and cyan rectangles.

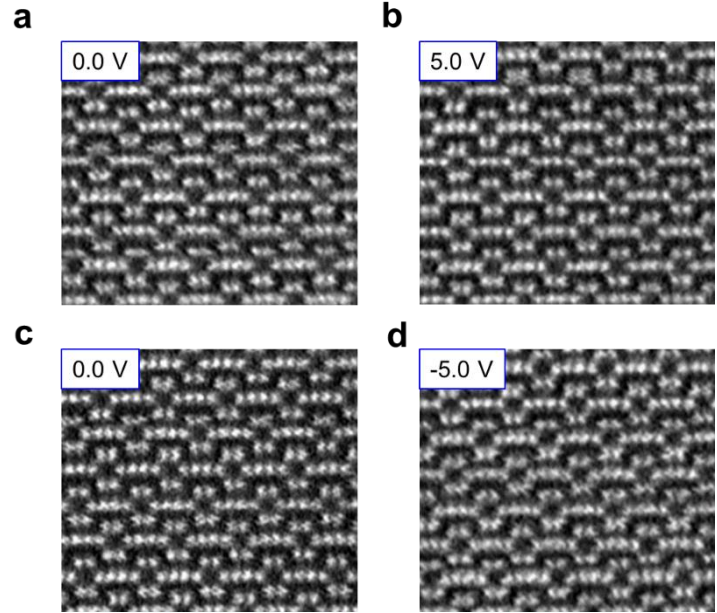

**Figure S4**. High-resolution HAADF-STEM images from an area inside a structural $D_1$ domain far away from a pre-existing non-polar domain wall. In this area, the application of ±5 V does not measurably displace Fe ions.

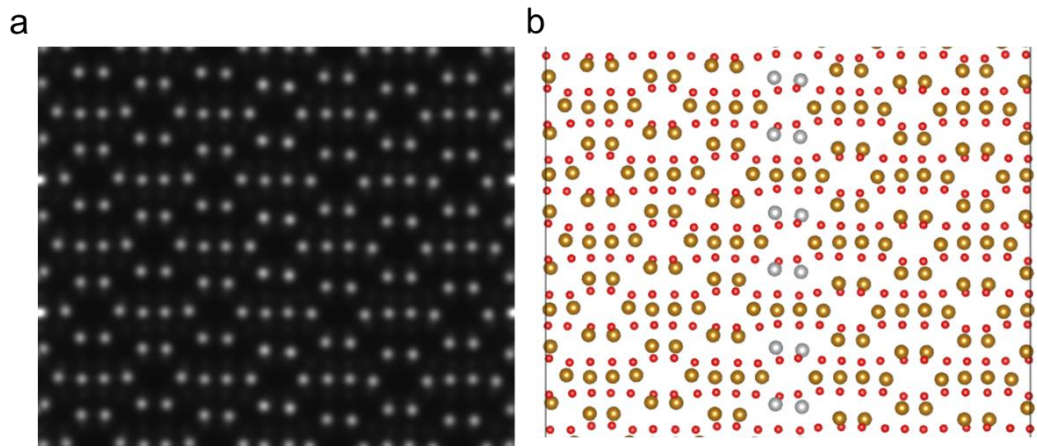

**Figure S5.** a) HAADF-STEM image simulation of the structural model shown in (b). The structural model in (b) is based on first-principles calculations. At the boundary between two polar domains, the smile and cry features disappear. The HAADF simulation of the structural model is performed by using Multislice calculation with xHREM. Aperture radius is 24 mrad. Third order and fifth order Cs are 0.005 mm and 0 mm, respectively. Inner and outer angles of dark field are 70 and 370 mrad, respectively.

**Static deformations and their influence on ferroelectric switching inside ε-Fe$_2$O$_3$ domains and around non-polar domain walls**

In this section we address the question how elastic deformations could influence ferroelectric switching in ε-Fe$_2$O$_3$. We calculate the piezoelectric tensor of ε-Fe$_2$O$_3$ (see Table 2 in the main manuscript) and the migration barriers in the deformed cells considering *Pbcn* as the reference non-polar structure (see Figure S6). We find the largest reduction of the energy barrier for a compressive in-plane lattice deformation, i.e., a lattice expansion along the polar axis. This observation could be used to design efficiently switchable strained ferroelectric films. Unfortunately, when the strain is purely a result of the piezoelectric effect, the reduction of the migration barrier is actually expected if the external electric field points along the polarization, i.e., in the most stable configuration. A reversal of the electric field will reduce the out-of-plane lattice parameter, which in turn leads to an increase of the switching barrier. Therefore, the piezoelectric effect with uniaxial strains does not support ferroelectric switching.

      We found that the shear deformation ε$_5$ lowers the energy barrier. This strain breaks the glide plane symmetry and lowers the symmetry of the transition state from *Pbcn* to *P*2/*c* as well as the symmetry of the polar structure from *Pna*2$_1$ to *Pc*, leading to a conventional group-subgroup relation in ferroelectric transitions (see Figure S6). Although the desired d$_{53}$ piezoelectric component for an out-of-plane electric field along the [001] polar axis is zero by symmetry, the ε$_5$ shear could be achieved via the d$_{51}$ piezoelectric component of an electric field applied along the [100] axis (see Figure S7). Therefore, one can assume that an electric field component along [100] axis could assist ferroelectric switching inside ε-Fe$_2$O$_3$ domains. This could be achieved by designing electric contacts on the thin films of ε-Fe$_2$O$_3$, leading to a gradient of electric field along [100] direction.

      Deformation ε$_4$ in *Pna*2$_1$, which is produced by the d$_{42}$ piezoelectric component if an electric field is applied along [010], enhances the migration barrier (see Figure S6). Moreover, the glide plane symmetry of the initial polar structure is not broken in this case. Therefore, the resulting piezoelectric tensor does not have a component that could induce the ε$_5$ shear deformation under an external electric field (see Figure S7).

      Static deformation ε$_6$ slightly decreases the migration barrier and breaks the glide plane symmetry, leading to *P*2$_1$ symmetry (see Figure S6). Unfortunately, this deformation cannot be achieved via the piezoelectric effect because all corresponding d$_{6i}$ components are zero in *Pna*2$_1$ (see Figure S7).

The presence of a vertical domain wall removes the glide plane symmetry orthogonal to the [100] axis. Actually, one can see the domain wall as a unit cell of the structure which is under $\varepsilon_6 = 5\%$ static deformation in half of the unit cell along the polar axis (i.e., for $0 \leq z \leq 0.5$) and under $\varepsilon_6 = -5\%$ in the other half of the unit cell. In general, large displacements of ions along the [100] direction in the region of the domain wall (see Figure 6g in the main manuscript) could be assigned to the $\varepsilon_6$ deformation. As in the case of the polar $\varepsilon$-$Fe_2O_3$ domain structure, this type of deformation leads to a local $P2_1$ space group. In contrast to $\varepsilon$-$Fe_2O_3$ domains where the unique axis is oriented along the polar axis, the unique axis is oriented along the [100] direction in $\varepsilon$-$Fe_2O_3$ with a vertical domain wall. Therefore, the $d_{61}$ piezoelectric tensor component (in coordinates of the standard crystallographic settings, see Figure S7) becomes active for perpendicular electric fields along [001] and could assist in domain wall motion by a monoclinic tilt of the polar *c* axis towards the *a* axis (i.e., via $\varepsilon_5$ shear in the *Pna*$2_1$ space group). Indeed, we found that because of asymmetric coupling of $\varepsilon_5$ strain with the ions in the domain wall region, which are situated above and below $z = 0.5$, the energy profile along the transition path becomes asymmetric, favoring the displacement of the domain wall in a particular direction (see Figure S6, bottom). We also considered the influence of static deformations $\varepsilon_4$ and $\varepsilon_6$ as well as axial deformations on the migration barrier of the domain wall and found that the barrier insignificantly increases under these types of strain.

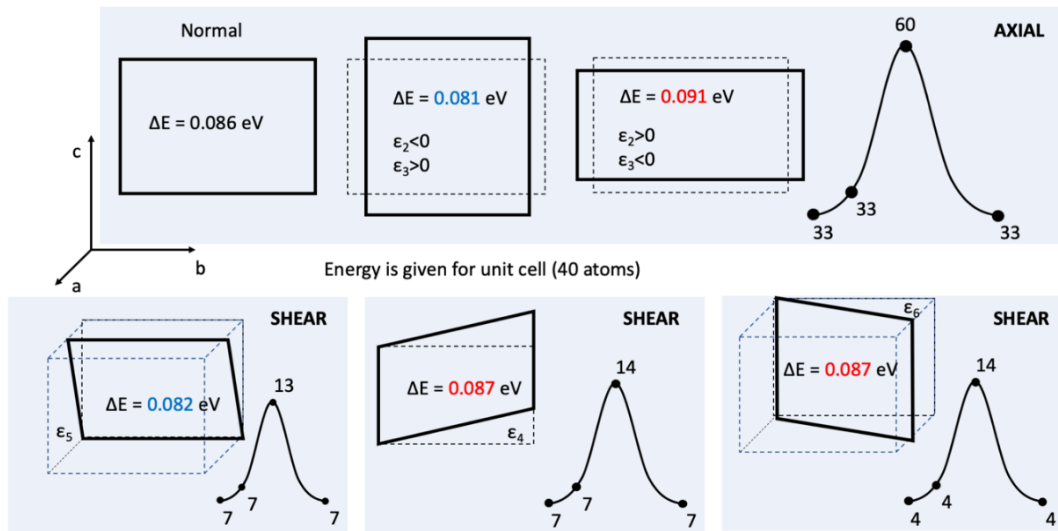
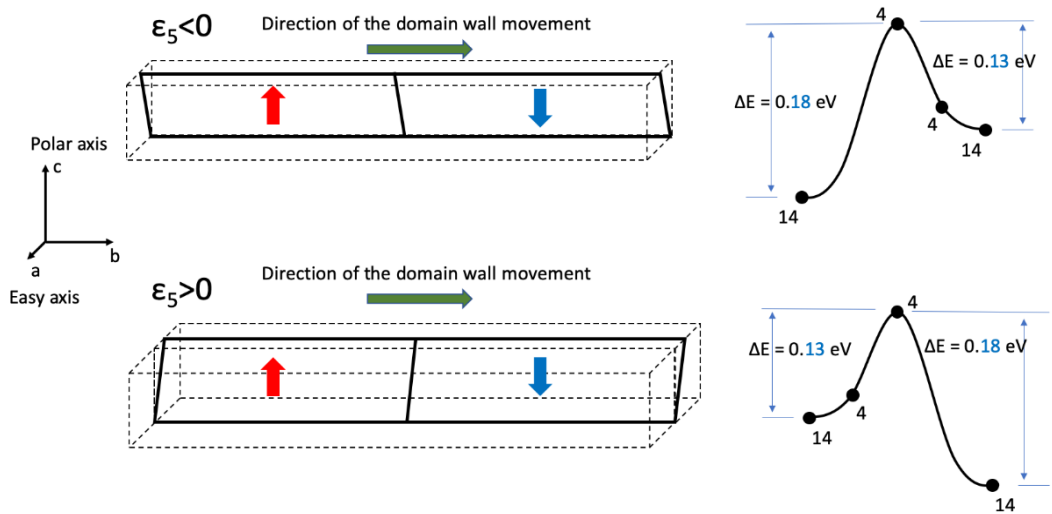

**Figure S6**. Up: Schematic presentation of the influence of axial and shear deformations ε = 1% on the migration barrier Δ$E$ inside a ε-$Fe_2O_3$ domain. The numbers in the schematic transition paths indicate space group numbers of the initial, transition, and final states along the migration paths. Down: Influence of $\varepsilon_5$ = 1% shear deformation on the migration barrier in multidomain ε-$Fe_2O_3$.

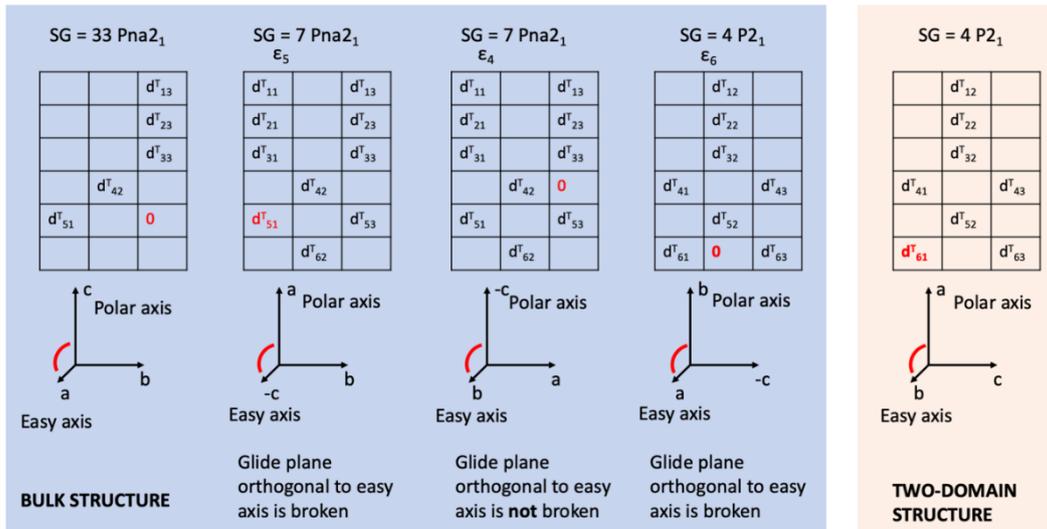

**Figure S7.** Piezoelectric tensors for different polar groups. The orientation of principal axes of the space group in standard setting with respect to the polar and easy axes of the *Pna*$2_1$ space group is schematically presented. Red arcs indicate the $\varepsilon_5$ shear distortion of the *Pna*$2_1$ space group. The external electric field is collinear to the polar axis.

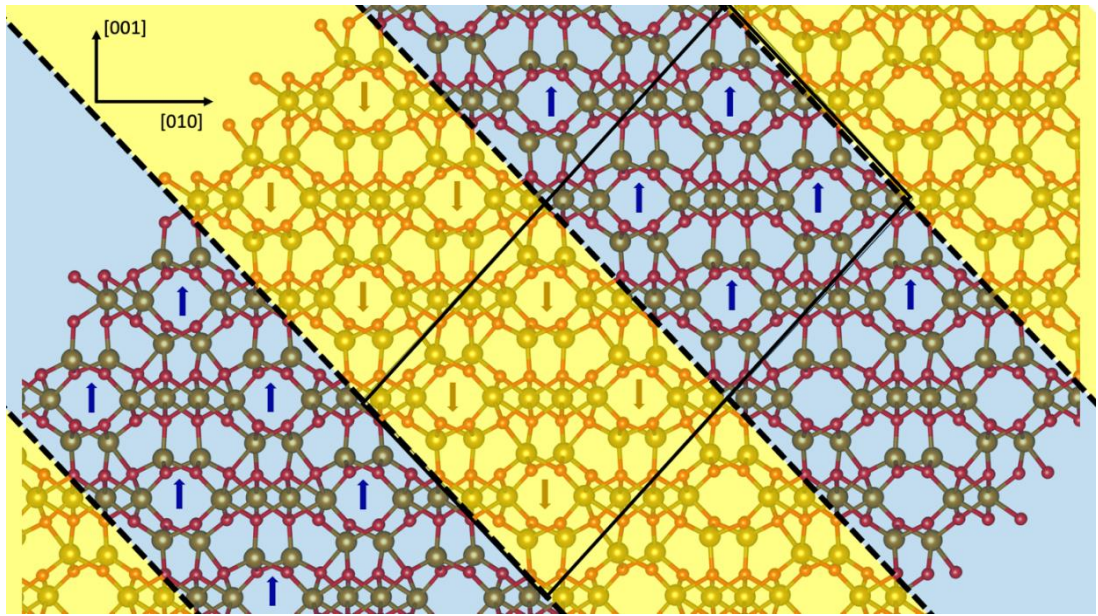

**Figure S8**. Model of inclined domain walls. The dashed lines indicate the positions of the domain walls. Arrows indicate the polarization direction in the ferroelectric domains.